 \documentclass[page-classic]{epl2} 

\title{Hawking temperature of Black Holes using the EUP and EGUP formalisms based on the Einstein-Bohr's Photon Box}
\shorttitle{Title} 

\author{N. Farahani\inst{1} \and H. Hassanabadi\inst{1,2} \and W.S. Chung\inst{3} \and B.C. L\" utf\"uo\u{g}lu\inst{2,4} \\ \and S. Zarrinkamar\inst{5}}
\shortauthor{N. Farahani \etal}

\institute{
  \inst{1} Faculty of Physics, Shahrood University of Technology, Shahrood, Iran.\\
  \inst{2} Department of Physics, University of Hradec Kr\'alov\'e,
				Rokitansk\'eho 62, 500 03 Hradec Kr\'alov\'e, Czechia. \\
  \inst{3} Department of Physics and Research Institute of Natural Science,
			College of Natural Science,
			Gyeongsang National University, Jinju 660-701, Korea. \\
  \inst{4} Department of Physics, Akdeniz University, Campus 07058 Antalya, Turkey. \\
  \inst{5} Department of Basic Sciences, Garmsar Branch, Islamic Azad University, Garmsar, Iran.
}

\abstract{
In this paper, by studying the Einstein-Bohr's photon box for weighting a photon, we find that the effective Newton constant can be proposed by the extended uncertainty principle and extended generalized uncertainty principle. We obtain the modified Hawking temperature, mass, specific heat, and entropy by using the modified Schwarzschild metric.}
\begin{document}

\maketitle

\section{Introduction}
{According to the ordinary quantum mechanics, there is a fundamental lower limit for the simultaneous measurement of certain pairs of physical quantities such as momentum and length. This limit is proportional to the Planck constant and derived from the Heisenberg algebra, therefore it is called the Heisenberg uncertainty principle  (HUP) \cite{Heisenberg}.  However, HUP does not predict a minimum uncertainty in position or momentum, solitary. On the other hand, the studies regarding the quantum gravity point to the existence of some additional terms not found in the HUP \cite{vagenas, vaagenas, Ali}. The predicted modification of Heisenberg algebra is associated with a parameter in the order of Planck length \cite{Planck} and it yields to derive a minimum uncertainty in position. This generalization of the HUP is named as the generalized uncertainty principle (GUP) \cite{GUP} and it is a well-known fact that its origin is based on the string theory \cite{string} and the Gedanken experiments \cite{Gedanken}. It is shown that in another generalization of the HUP,  namely in the extended uncertainty principle (EUP), a minimum  uncertainty in momentum can be obtained associated with an extension parameter that is proportional to the cosmological constant \cite{EUP}. Meanwhile, in another work, it is shown that if the (anti)-de Sitter ((A)dS) background is taken into account instead of the Minkowski space-time,  then HUP is needed to be extended with an additional term that is inversely proportional to the square of the (A)dS radius \cite{8,9}. It is worth noting  that the GUP and EUP are thought to play an important role in the early and later stages of the universe, respectively \cite{UUnruh}. Therefore, a combination of these two generalized forms, hereafter the extended generalized uncertainty principle (EGUP), is formulated to consider the minimal uncertainties together in position and momentum in various studies \cite{Haw-EGUP, EGUP2}.

In 1930, Bohr and Einstein, two of the greatest physicists in history drew swords to each other on a very fundamental property of quantum mechanics, HUP \cite{Bohr, Torre, Hnizdo, Dieks}. Einstein devised a Gedanken experiment, namely the photon-box experiment, to show up an inconsistency in the time-energy uncertainty relation. After one long and sleepless night, Bohr defended the HUP and saved his career \cite{E-B}. In this letter, we revisit the photon-box experiment by considering the EUP and EGUP formalisms. We aim to realize how the gravitational field strength and correspondingly, the Newton constant would be modified based on the EUP and EGUP. Next, we want to derive the Hawking temperature \cite{Hawking} and altered thermodynamic properties of the black hole, taking into account the effective (modified) Newton constant in a Schwarzschild black hole \cite{HOVG, HOVG1, HOVG2,B}.

We construct this letter as follows: In the following section, we briefly discuss the Einstein-Bohr debate. Then, in the third and fourth sections, by introducing EUP and EGUP in AdS space-time, we derive the effective gravitational field strength, thus, the modified Schwarzschild black holes metric. After that, we discuss the modified Hawking temperature, the mass-temperature, specific heat-temperature and entropy-temperature relations for the EUP and EGUP black holes based on Einstein-Bohr's photon box Gedanken experiment, respectively. In the conclusion, we compare the modified Hawking temperature for EUP, GUP, and EGUP approaches.}

\section{Einstein-Bohr’s photon box}
{Einstein's purpose in designing this Gedanken experiment was to reject the HUP, explicitly. He considered an experiment box with a clock inside and a clock-related gap \cite{EE-BB}. The gap associated with the clock is set against another gap at a specified interval. At a time interval, the light pulse passes through these two gaps. If this light pulse is located at a farther distance, at a certain distance from the gap, it will be detected and its energy will be measured with arbitrary precision. Measuring energy and time with arbitrary precision would have meant the collapse of Heisenberg's energy-time relationship \cite{E2}.

Heisenberg believed in successive measurements of the position and momentum. However, the uncertainty of the momentum or position of a particle is related to the position or momentum that is measured at that moment. Hence,  the HUP  is not valid for retroactive measurements. If we want to turn this retroactive measurement into a predictive one, we have to consider the energy exchange during the gap movement. Let us assume that the gap moves at the speed of $v$. Then, the uncertainty of the momentum $\Delta p$ during the radiation of the photon has to result in the uncertainty of $\Delta E=v \Delta p$. Since this energy exchange is executed in the constant gap width, $d$, then the momentum uncertainty has at least to be equal to $\frac{\hbar}{d}$. Therefore, we get}
\begin{eqnarray}
\Delta E \geq \frac{\hbar v}{d}.
\end{eqnarray}
{To increase the accuracy of energy determination,  $\frac{v}{d}$ ratio has to be reduced as much as possible either by decreasing $v$ or by increasing $d$. In both cases, the precision of the timing is disturbed, thereby, according to $\Delta t \approx \frac{d}{v}$ we arrive at \cite{EE-BB} }
\begin{eqnarray}\label{Et}
\Delta E \Delta t \geq \hbar.
\end{eqnarray}
{Einstein developed another Gedanken experiment scheme that will be named as the "Clock in the Box" experiment. He considered a box that contains electromagnetic radiation with full reflective inner walls. He assumed one of the walls has a shutter on and it is opening and closing by a clock-related mechanism inside the box. The setting is such that the slot opens at time $t=t_{0}$ for a short and arbitrary time interval, i.e. $t_{2}-t_{1}$, thereby a photon can be emitted. By weighing the mass of the box just before and after the photon emission, it is possible to determine the energy difference of the box with the least error $ \Delta E $ regarding the mass-energy relation $E=m c^2$. According to the energy conservation law, the energy difference is related to the energy of the emitted photon. Therefore, the energy of the photon and the time that it needs to reach the plate from the device can be predicted by arbitrary uncertainty $\Delta E$ and $\Delta t$. This is, of course, contrary to the Heisenberg uncertainty relation.

However, after just one night, Bohr came up with a solution that demonstrated that the HUP remains valid. His defense was based on Einstein's general theory of relativity. Einstein did not take the time-dilation effect into account. One can prove the  validity of HUP as follows: During the emission of the photon, the uncertainty in the momentum of the box is $\Delta p= \Delta(m v)$ where $v=gt$. Here, $t$ is the time for the observer outside the box. Then, we employ them in the HUP. We find}
\begin{eqnarray}\label{m}
\Delta X \geq \frac{\hbar}{\Delta m g t}.
\end{eqnarray}
{Since there is a time dilation,  the observer in the box (clock) experiences a time uncertainty, $\Delta t$. In \cite{Bohm}, it is given in terms of  the vertical
position uncertainty as follows:}
\begin{eqnarray}\label{t}
\Delta t=\frac{g \Delta X}{c^2}t.
\end{eqnarray}
By substituting $t$ from eq.~(\ref{t}) into eq.~(\ref{m}) we arrive at
\begin{eqnarray}
\Delta t \geq \frac{\hbar}{\Delta m c^2}.
\end{eqnarray}
Following the mass-energy relation we can express the conventional HUP as given in eq.~(\ref{Et}). On the other hand, by substituting eq.~(\ref{t})  into eq.~(\ref{Et}), we find \cite{energy-time}
\begin{eqnarray}\label{EE}
\Delta E \geq \frac{c^2 \hbar}{g t \Delta X}.
\end{eqnarray}

\section{Einstein-Bohr’s photon box based on the EUP}
{In this section we consider Bohr's argument throughout the EUP. As far as we know, Xiang \ emp {et al.} recently used the GUP instead of the HUP and derived an effective Newtonian constant \cite {file}. We believe that the use of EUP within the Bohr's argument is a very interesting open problem, as Mignemi reported that the EUP correction can be derived from the geometry of the (A)dS spacetime \cite{EGUP2}. The EUP is based on the $\alpha$ parameter, that is defined as $3 \alpha^2=|\Lambda|=10^{-52}m^{-2}$ where $\alpha^2=\frac{1}{L^2_{H}}$. Here,  $L_{H}$ is the radius of the (A)dS spacetime and $\Lambda$ is the cosmological constant  \cite{de sitter, EUP2}. It is worth noting that in quantum field theory, dS spacetime is defined in positively curved spacetime with positive cosmological constant and radius, while AdS spacetime is defined in negatively curved spacetime with negative cosmological constant and radius \cite{AdS}. We start with the following modified commutation relation}
\begin{eqnarray}\label{EUP}
\left[ X,P\right] =i \hbar (1\pm \alpha^2 X^2),
\end{eqnarray}
where $\alpha^2=\frac{\alpha_0 |\Lambda|}{3}$ while $\alpha_0$ has the order of unity and dimensionless. Then, the EUP appears as \cite{NEUP, NNEUP}
\begin{eqnarray}
\Delta X \Delta P \geq \frac{\hbar}{2} (1+\frac{\alpha_{0}|\Lambda|}{3} \Delta X^2).
\end{eqnarray}
However, we can employ a more general form of the modified commutation relation
\begin{eqnarray}
\left[ X,P\right] =i \hbar \xi(X),~~~~~~\xi(X)=\left(1+ \frac{\alpha_0 |\Lambda|}{3} X^2\right),
\end{eqnarray}
hereafter, $\xi (X) \equiv \xi $.

\begin{figure}
\onefigure[scale=0.6]{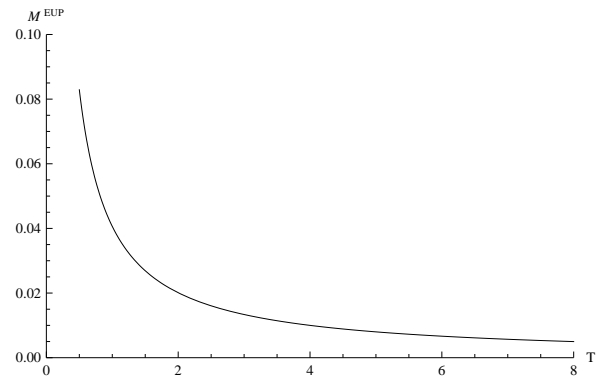}
\caption{Plot of the modified mass-temperature relation for the $M^{EUP}$ with $\alpha_{0}|\Lambda|=0.5$ where we set $\hbar = c = k_B = G= 8 \pi= 1$.}
\label{fig.MEUP}
\end{figure}

\begin{figure}
\onefigure[scale=0.6]{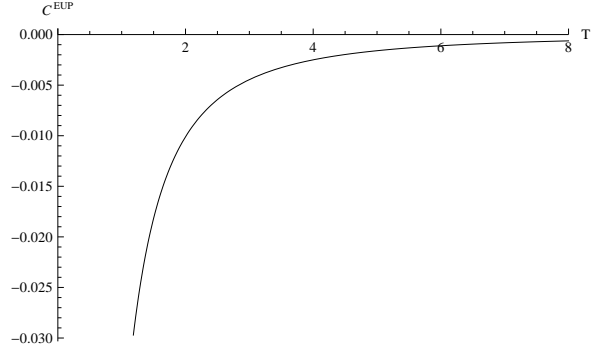}
\caption{Plot of the modified specific heat-temperature relation for the $C^{EUP}$ with $\alpha_{0}|\Lambda|=0.5$  where we set $\hbar = c = k_B = G= 8 \pi= 1$.}
\label{fig.CEUP}
\end{figure}

\begin{figure}
\onefigure[scale=0.6]{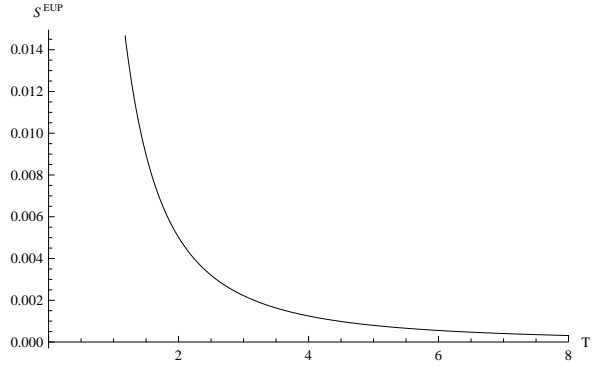}
\caption{Plot of the modified entropy-temperature relation for the $S^{EUP}$ with $\alpha_{0}|\Lambda|=0.5$ where we set $\hbar = c = k_B = G= 8 \pi= 1$}
\label{fig.SEUP}
\end{figure}

{In the Gedanken experiment, after the photon is released, the box moves upward as it becomes lighter. In order to lower the box to its previous level the lowest and quantum-mechanically meaningful mass is added to the system. It corresponds to the released photon's weight, namely $g \Delta m$.
This procedure is supposed to be executed in a period of time $t$, \cite {file}. Thus, we arrive at}
\begin{eqnarray}
\frac{\Delta P_{min}}{t}=\frac{\xi \hbar}{t \Delta X} \leq g \Delta m, \label{EUPcan}
\end{eqnarray}
or we can write
\begin{eqnarray}
\xi \hbar=\Delta X \Delta P_{min} \leq \Delta m (g \Delta X)t. \label{eq11}
\end{eqnarray}
Then, by substituting eq.~(\ref{t}) into eq.~(\ref{eq11}), we get
\begin{eqnarray}\label{mm}
\xi \hbar \leq c^2 \Delta m \Delta t=\Delta E \Delta t.
\end{eqnarray}
{We employ $\Delta t$ from eq.~(\ref{t}) in the above inequality.  We obtain the uncertainty in the energy of the released photon as follows}
\begin{eqnarray}
\Delta E^{EUP}\geq \frac{\xi \hbar c^2}{g \Delta X t}=\frac{ \hbar c^2}{g^{\prime} \Delta X t}.
\end{eqnarray}
{When we compare our result with equation eq.~(\ref{EE}), we observe the only difference as the replacement of $g^{\prime}$ at $g$. Thus, we conclude that in the EUP formalism  the modified gravity of Earth  becomes}
\begin{eqnarray}
g^{\prime}=\frac{g}{\xi}=\frac{g}{1+\frac{\alpha_{0} |\Lambda|}{3} X^2}.
\end{eqnarray}
Note that when $\alpha=0$, we find the HUP limit. Then, we express the effective gravitational field strength from the well-known definition $g=\frac{GM}{R^2}$ \cite{G}.
\begin{eqnarray}
G^{\prime}=\frac{G}{1+\frac{\alpha_{0} |\Lambda|}{3} X^2}.
\end{eqnarray}
{Therefore, we conclude that we come up with a curved space from a  modified effective Newton constant.Next, as done in \cite{file}, we change $G$ with $G^{\prime}$ and examine a new Schwarzschild metric with the form \cite{metric}}
\begin{eqnarray}\label{metric}
ds^2=-\left(1-\frac{2G M}{\xi c^2 r}\right)c^2 dt^2+\left(1-\frac{2G M}{\xi c^2 r}\right)^{-1}dr^2+r^2 d\Omega^2.
\end{eqnarray}
{We explore solutions for the modified Schwarzschild black hole in which the mass is taken as $\tilde{M}=\frac{M}{\xi(X)}$. We obtain the Hawking temperature, \cite{Haw}, as follows:}
\begin{eqnarray}\label{Hawking T}
T_{H}=\frac{\hbar c^3}{8 \pi G k_{B} \tilde{M}}.
\end{eqnarray}
On the other hand, we consider $X$ of order $\Delta X$. Close to the black hole  the uncertainty is proportional to the black hole event horizon \cite{Adler}, thus, we write $\Delta X=\eta r_{s}$, where $r_{s}=\frac{2GM}{c^2}$. We find
\begin{eqnarray}
T=\frac{\hbar c^3}{8 \pi G k_{B} M}\left(1+\frac{4}{3}\frac{\alpha_{0} |\Lambda| G^2 M^2 \eta^2}{c^2}\right).
\end{eqnarray}
We solve the above equation while considering $\eta=2 \pi$,  we obtain
\begin{eqnarray}\label{mass E}
M^{EUP}=\frac{3 k_{B} T c}{4 \alpha_{0}|\Lambda| G \pi \hbar} \left(1 \pm \sqrt{1-\frac{1}{3}\frac{\alpha_{0} |\Lambda|c^2 \hbar^2}{k_{B}^2 T^2}}\right),
\end{eqnarray}
which implies a minimum temperature as
\begin{eqnarray}
T\geq  T^{EUP}_{min}=\frac{\hbar c}{k_{B}}\sqrt{\frac{\alpha_0 |\Lambda|}{3}}.\label{eq20}
\end{eqnarray}
{We employ  eq.~(\ref{eq20}) in eq.~(\ref{mass E}). In order to obtain the minimum mass for EUP, we use the negative quantity. We find}
\begin{eqnarray}
M^{EUP}_{min}=\frac{3 k_{B} T c}{4 \alpha_{0}|\Lambda| G \pi \hbar}. \label{EUPMmin}
\end{eqnarray}
{Considering eq.~(\ref{mass E}) for small values of $\alpha_{0} |\Lambda|$  with a Taylor expansion, we find}
\begin{eqnarray}
M^{EUP}=\frac{\hbar c^3}{8 \pi G k_{B}}\left(\frac{1}{T}+\frac{\alpha_0 |\Lambda|\hbar^2 c^2}{12 k^2_{B} T^3}+\frac{\alpha^2_0 \Lambda^2 c^4 \hbar^4}{72 k^4_{B} T^5}+...\right).
\end{eqnarray}
{It is worth noting that, for $\alpha_0=0$, we recover the ordinary case result,  $M=\frac{\hbar c^3}{8 \pi G k_{B} T}$. Then, we demonstrate the mass function versus temperature in Fig.~(\ref{fig.MEUP}).
Next, we examine the specific heat of the EUP black hole which can be derived out of eq.~(22) as done in  \cite{Haww, Unruh}. We find}
\begin{eqnarray}
C^{EUP}&=&c^2 \frac{dM^{EUP}}{dT},\\
&=&-\frac{\hbar c^5}{8 \pi G k_{B}}\left(\frac{1}{T^2}+\frac{\alpha_0 |\Lambda| \hbar^2 c^2}{4 k^2 T^4}+\frac{5\alpha^2_0 \Lambda^2 \hbar^4 c^4}{72 k^4_B T^6}+...\right).\nonumber 
\end{eqnarray}
{We observe that the specific heat in the EUP black hole always is negative. We depict the specific heat versus temperature in Fig.~(\ref{fig.CEUP}) to present its characteristic behavior. Next, we derive entropy for the EUP black hole by integrating eq.~(23).  We find}
\begin{eqnarray}
S^{EUP}&=&\int \frac{C^{EUP}}{T} dT, \\
       &=& \frac{\hbar c^5}{8 \pi G k_{B}}\left(\frac{1}{2 T^2}+\frac{\alpha_0 |\Lambda| \hbar^2 c^2}{16 k^2_{B} T^4}+\frac{5 \alpha^2_0 \Lambda^2 \hbar^4 c^4}{432 k^4_{B} T^6}+...\right).\nonumber
\end{eqnarray}
{In Fig.~(\ref{fig.SEUP}), we plot the modified entropy-temperature function versus temperature. Finally, by taking the negative sign and expanding eq.~(\ref{mass E}), we obtain the modified Hawking temperature based on EUP in terms of  temperature.}
\begin{eqnarray}\label{T_{EUP}}
T^{EUP}_{H}=T_{H}\left(1+\frac{\alpha_{0}|\Lambda| \hbar^2 c^2}{12 k_{B}^2 T^2}\right).
\end{eqnarray}
{In the following section, by following the same strategy we are going to obtain the EGUP formalism equivalents of the functions we have obtained in the EUP formalism.}

\section{Einstein-Bohr’s photon box based on the  EGUP}
\begin{figure}
\onefigure[scale=0.6]{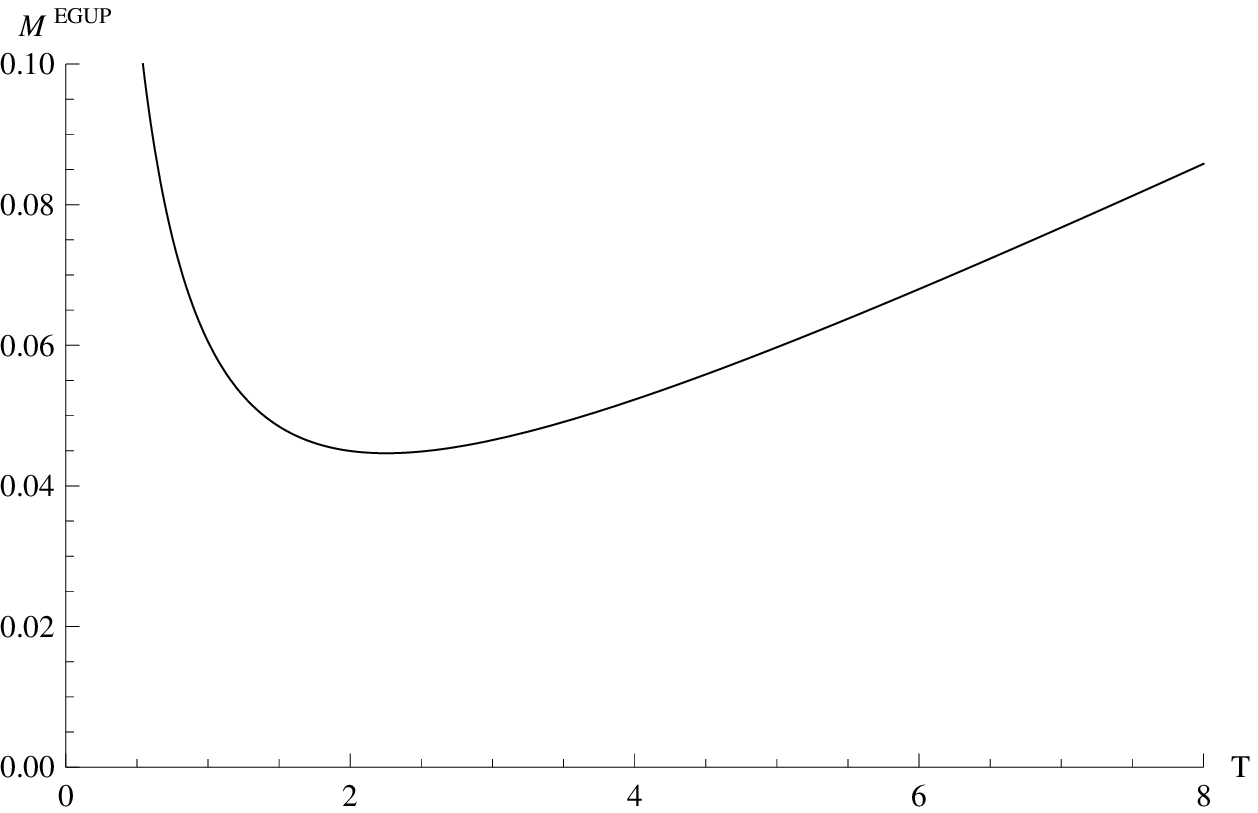}
\caption{Plot of the modified mass-temperature relation for the $M^{EGUP}$ with $\alpha_{0}|\Lambda|=0.5$ and $\beta_0 \ell^2_{P}=0.5$ where we set $\hbar = c = k_B = G= 8 \pi= 1$.}
\label{fig.MEGUP}
\end{figure}

\begin{figure}
\onefigure[scale=0.6]{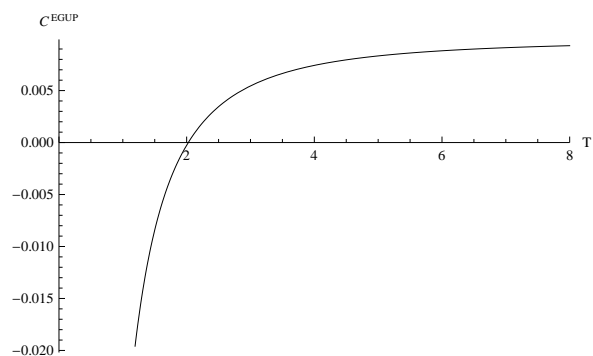}
\caption{Plot of the modified specific heat-temperature relation for the $C^{EGUP}$ with $\alpha_{0}|\Lambda|=0.5$ and $\beta_0 \ell^2_{P}=0.5$ where we set $\hbar = c = k_B = G= 8 \pi= 1$.}
\label{fig.CEGUP}
\end{figure}

\begin{figure}
\onefigure[scale=0.6]{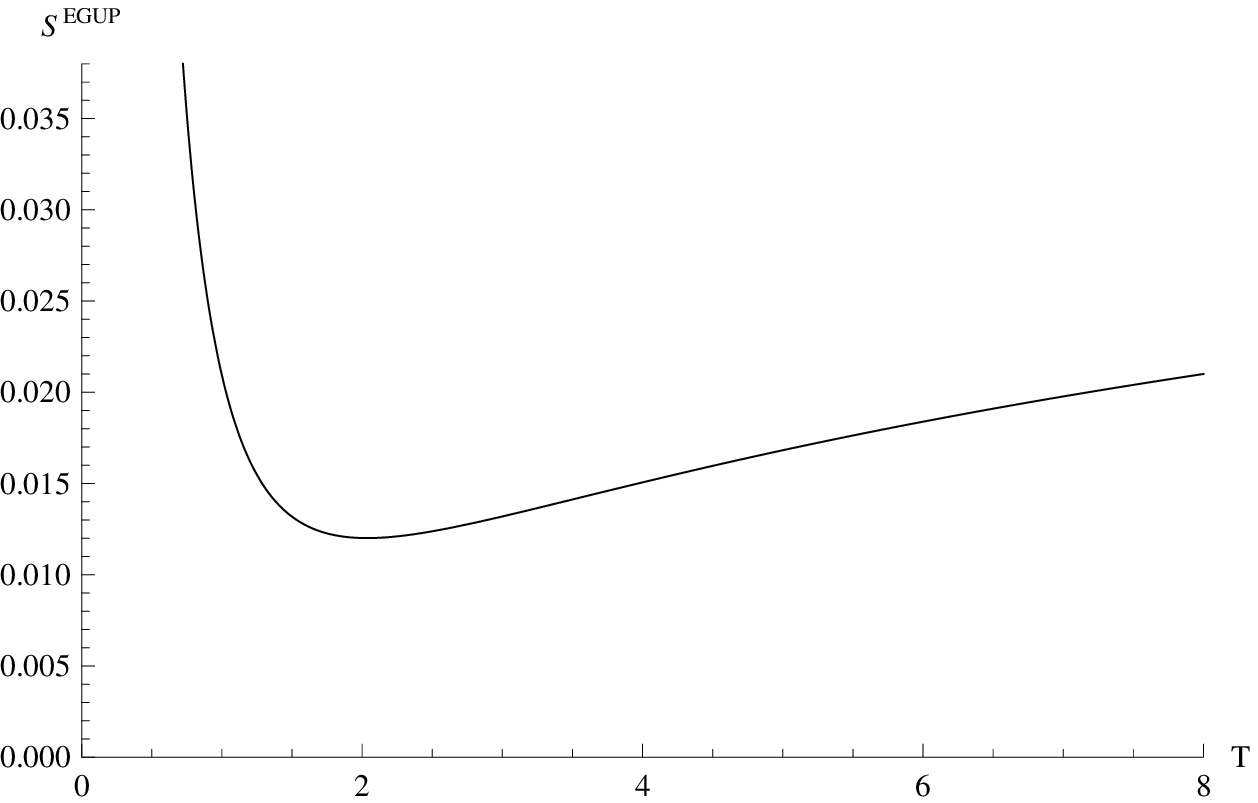}
\caption{Plot of the modified entropy-temperature relation for the $S^{EGUP}$ with $\alpha_{0}|\Lambda|=0.5$ and $\beta_0 \ell^2_{P}=0.5$ where we set $\hbar = c = k_B = G= 8 \pi= 1$.}
\label{fig.SEGUP}
\end{figure}

\begin{figure}
\onefigure[scale=0.7]{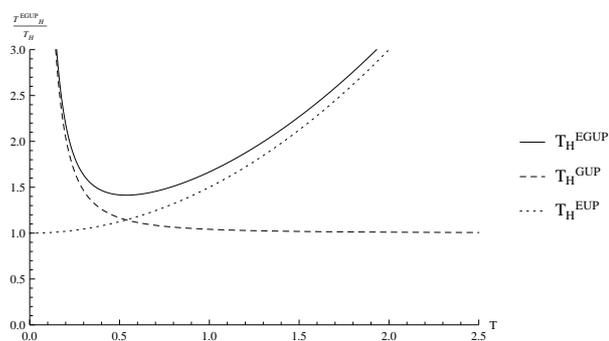}
\caption{Plot of the modified Hawking temperature-temperature relation for the $T^{EGUP}_{H}$, $T^{EUP}_{H}$ and $T^{GUP}_{H}$ with $\beta_{0}\ell^2_{P} = 0.5$ and $\alpha_{0}|\Lambda|=0.5$ where we set $\hbar = c = k_B = G= 8 \pi= 1$.}
\label{fig.T}
\end{figure}

{We start by introducing EGUP as a linear  combination of the EUP and GUP \cite{EGUPP, EGGUP}}
\begin{eqnarray}\label{EGUP}
\Delta X \Delta P \geq \frac{\hbar}{2} \left(1+\frac{\alpha_{0}|\Lambda|}{3} \Delta X^2 +\frac{\beta_{0}\ell^2_{P}}{\hbar^2} \Delta P^2\right),
\end{eqnarray}
{which can be obtained from the following modified commutation relation}
\begin{eqnarray}
\left[ X,P\right] =i \hbar \left(1+ \frac{\alpha_{0}|\Lambda|}{3} X^2+\frac{\beta_{0}\ell^2_{P}}{\hbar^2} P^2\right).
\end{eqnarray}
{Here, $\ell_{P}$ is the Planck length and $\beta_0$ is the parameter of the order of unity. Then, we rewrite the modified commutation relation in the most general form, $\left[ X,P\right] =i \hbar \zeta$, where $\zeta=\zeta(X,P)$. We follow the same arguments of the previous sections and express}
\begin{eqnarray}
\frac{\Delta P_{min}}{t}=\frac{\zeta \hbar}{t \Delta X} \leq g \Delta m,
\end{eqnarray}
{instead of eq.~(\ref{EUPcan}) \cite{file, B}. Alternatively, we can write}
\begin{eqnarray}
\zeta \hbar=\Delta X \Delta P_{min} \leq \Delta m (g \Delta X)t.
\end{eqnarray}
{By replacing eq.~(\ref{t}) into eq.~(29), we find}
\begin{eqnarray}
\zeta \hbar \leq c^2 \Delta m \Delta t=\Delta E \Delta t.
\end{eqnarray}
{We observe that the uncertainty in the energy of the released photon is restricted by}
\begin{eqnarray}
\Delta E^{EGUP}\geq \frac{\zeta \hbar c^2}{g \Delta X t}=\frac{ \hbar c^2}{g^{\prime} \Delta X t}.
\end{eqnarray}
{Here, we use $g^{\prime}=\frac{g}{\zeta}$. Then, as we have done in the previous section, we find the effective gravitational field strength in the form of}
\begin{eqnarray}
G^{\prime}&=&\frac{G}{\left(1+ \frac{\alpha_{0}|\Lambda|}{3} X^2+\frac{\beta_{0}\ell^2_{P}}{\hbar^2} P^2\right)}.
\end{eqnarray}
{Next, we use the EGUP effective gravitational field strength to examine the modified Schwarzschild black hole while $M^{\prime}=\frac{M}{\zeta}$. Then, eq.~(\ref{Hawking T}) changes its form to}
\begin{eqnarray}
T=\frac{\hbar c^3}{8 \pi G k_{B} M} \left(1+ \frac{\alpha_{0}|\Lambda|}{3} X^2+\frac{\beta_{0}\ell^2_{P}}{\hbar^2} P^2\right).
\end{eqnarray}
{After that according to \cite{ling1}, we assume $P=\frac{k_{B}T}{c}$ and $X=\frac{2Gm \eta}{c^2}$. We find}
\begin{eqnarray}
M=\frac{3 k_{B} c T}{4 \alpha_{0} |\Lambda| \hbar G \pi}\left(1 \pm \sqrt{1-\frac{\alpha_{0}|\Lambda| \beta_{0}\ell^2_{P}}{3} -\frac{\alpha_{0} |\Lambda| c^2 \hbar^2}{3 k^2_{B} T^2}}\right).
\label{eq34}
\end{eqnarray}
{In Fig.~(\ref{fig.MEGUP}), we plot the mass-temperature function versus temperature. Then, we explore the temperature expression that corresponds to the minimum mass value of the black hole. We get	}
\begin{eqnarray}
T\geq T^{EGUP}_{0}=\frac{\hbar c}{k_{B}}\sqrt{\frac{\alpha_0 |\Lambda|}{3\left(1-\frac{\alpha_0 |\Lambda| \beta_0 \ell^2_P}{3}\right)}}.\label{eq35}
\end{eqnarray}
{We substitute eq.~(\ref{eq35}) into eq.~(\ref{eq34}) and we use the positive quantity. We obtain the minimum mass value as}
\begin{eqnarray}
	M\geq M^{EGUP}_{min}= \frac{3 k_{B} T c}{4\pi \alpha_{0} |\Lambda| \hbar G }. \label{EGUPMmin}
\end{eqnarray}
{We compare minimal mass values of the EUP and EGUP black holes by matching eqs.~(\ref{EUPMmin}) and (\ref{EGUPMmin}). Notably, we observe that $M^{EGUP}_{min}= M^{EUP}_{min}$. For the small values of  $\alpha_0 |\Lambda|$ and $\beta_0 \ell^2_P$ parameters we Taylor expand eq.~(\ref{eq34}) by considering the negative sign. We find } \cite{UUnruh}
\begin{eqnarray}
	M^{EGUP}&=&\frac{\hbar c^3}{8 \pi G k_{B}}\bigg(\frac{1}{T}+\frac{\alpha_0 |\Lambda|\hbar^2 c^2}{12 k^2_{B} T^3}\nonumber \\
&+&\frac{\beta_0 \ell^2_{P} k_{B}^2 T}{c^2}+\frac{\alpha_0 |\Lambda| \beta_0 \ell^2_{P} \hbar^2}{6 T}\bigg).
	\end{eqnarray}
{Then, we derive the specific heat function of the EGUP black hole case out of eq.~(37) \cite{UUnruh}}
\begin{eqnarray}
C^{EGUP}&=&c^2 \frac{dM^{EGUP}}{dT}=\frac{\hbar c^5}{8 \pi G k_{B}}\bigg(-\frac{1}{T^2}+\frac{\beta_0 \ell^2_P k^2_{B}}{c^2} \nonumber \\
&-&\frac{\alpha_0 |\Lambda| \hbar^2 c^2}{4 k^2_{B}T^4}-\frac{\alpha_0 |\Lambda| \beta_0 \ell^2_P \hbar^2}{6 T^2}\bigg).
\end{eqnarray}
{In Fig.~(\ref{fig.CEGUP}), we plot the modified specific heat function versus temperature. We observe that the specific heat function is equal to zero at a particular temperature value. We find this temperature from the following expression}
\begin{eqnarray}
	T_0=\frac{\hbar c}{k_B \beta_0 \ell_P^2}\sqrt{\frac{\left(6+(\alpha_0 |\Lambda|  \beta_0 \ell^2_P \hbar)^2\right)}{12 \hbar^2}}\nonumber \\
\times\Bigg[1+\sqrt{1+\frac{36(\alpha_0 |\Lambda|  \beta_0 \ell^2_P \hbar)^2}{(6+(\alpha_0 |\Lambda|  \beta_0 \ell^2_P \hbar)^2)^2}}\Bigg]^{\frac{1}{2}}.
\end{eqnarray}
{Next, we derive the entropy function for the  EGUP black holes out of eq.~(38). We find}
\begin{eqnarray}
S^{EGUP}&=&\int \frac{C^{EGUP}}{T} dT, \nonumber\\
&=&\frac{\hbar c^5}{8 \pi G k_{B}}\bigg(\frac{1}{2T^2}+\frac{k^2_B \beta_0 \ell^2_P }{c^2}\ln T \nonumber\\
&+&\frac{\alpha_0 |\Lambda| \hbar^2 c^2}{16 k^2_{B} T^4}+\frac{\alpha_0 |\Lambda| \beta_0 \ell^2_P \hbar^2}{12 T^2}\bigg).
 \end{eqnarray}
{In Fig.~(\ref{fig.SEGUP}), we depict the modified entropy-temperature function versus temperature. We observe a minimal value that corresponds to $T_0$ temperature that is derived in eq.~(39). Note that at this temperature modified specific heat vanishes as well. Finally, we expand eq.~(34) by considering the negative sign. We obtain}
\begin{eqnarray}\label{T_{EGUP}}
\frac{T^{EGUP}_{H}}{T_{H}}=\bigg(1+\frac{\alpha_{0} |\Lambda| c^2 \hbar^2}{12 k^2_B T^2}+\frac{\beta_{0}\ell^2_{P} k^2_B T^2}{\hbar^2 c^2}+\frac{\alpha_{0} |\Lambda| \beta_{0}\ell^2_{P}}{6}\bigg).
\end{eqnarray}
{Finally, in Fig.~(\ref{fig.T}), we plot $T^{EGUP}_{H}$, $T^{EUP}_{H}$ and $T^{GUP}_{H}$  versus temperature. Here, we take the form of $T^{GUP}_{H}$ function from the reference  \cite{file}. We see that $T^{EGUP}_{H}$ and $T^{EUP}_{H}$ have similar
behaviors at high temperatures, i.e. $T^{EGUP}_{H}$  and $T^{EUP}_{H}$ are greater than $T_{H}$. We also note that $T^{EGUP}_{H}$ and $T^{GUP}_{H}$ have similar behavior at low temperatures, i.e. $T^{EGUP}_{H}$ and $T^{GUP}_{H}$ are greater than $T_{H}$. However, $T^{EUP}_{H}$ and $T^{GUP}_{H}$ don't present a similar behavior and they coincide at the temperature \cite{Haw-EGUP, UUnruh}.}
\begin{eqnarray}
T^{H}_0=\frac{\hbar c}{k \sqrt{2}} \left(\frac{\alpha}{3 \beta}\right)^{1/4}. \label{son}
\end{eqnarray}

\section{Conclusion}
{In this paper, we studied the Hawking temperature and the modified thermodynamic properties for the Schwarzschild black hole in the EUP and EGUP approaches. As shown in eqs. (14), (15) and (32), the effective gravitational field and the corresponding Newton constant are modified in the presence of EUP and EGUP. This result allows us to investigate the modified Hawking temperature and the corresponding thermodynamic properties for the black hole. Then, by obtaining the modified Hawking temperature in the presence of EGUP, we obtained the remarkable consequence that its value was the sum of the modified Hawking temperature in the presence of EUP and GUP, and an additional term which had both of generalization parameters $(\beta_0 \ell^2_P \alpha_0 \left| \Lambda \right| )$. This is the result we obtained after studying the mass, specific heat, and entropy of the Schwarzschild black hole. This result indicates that the modified Hawking temperature obtained for the Schwarzschild black hole has a higher generality , which is clearly shown in Fig.~(7). As can be seen, the modified Hawking temperatures of EGUP and GUP behave similarly at low temperatures. As the modified Hawking temperatures of EGUP and EUP behave similarly at high temperatures. On the other hand, it can be pointed out that the modified Hawking temperatures do not resemble a similar behavior in GUP and EUP cases, they only coincide at the temperature given in eq.~(\ref{son}). While the modified Hawking temperature in the presence of EGUP  gives us a more accurate answer. Also, after examining and plotting the specific heat and entropy of the Schwarzschild black hole in the presence of EGUP, we conclude that the entropy reaches its lowest value at the temperature $T_{0}$ in eq.~(39), and this is the temperature at which the specific heat is zero. Notably, we found that $M^{EGUP}_{min}= M^{EUP}_{min}$ for small values $\alpha_0 |\Lambda|$ and $\beta_0 \ell^2_P$.}

\acknowledgments
{The authors thank the referee for a thorough reading of our manuscript and for constructive suggestion.} B. C. L\" utf\"uo\u{g}lu, was partially supported by the Turkish Science and Research Council (T\"{U}B\.{I}TAK).

\end{document}